\documentstyle[12pt]{article}
\input tcilatex
\begin{document}

\date{}
\title{Exact Vacuum Solutions of Jordan, Brans-Dicke Field Equations}
\author{Sergey Kozyrev\thanks{
e-mail : {\ Sergey@tnpko.ru}} \\
%EndAName
\\
Scientific center gravity wave studies "Dulkyn".}
\maketitle

\begin{abstract}
We present the static spherically symmetric vacuum solutions of the Jordan,
Brans-Dicke field equations. The new solutions are obtained by considering a
polar Gaussian, isothermal and radial hyperbolic metrics.
\end{abstract}


\section{Introduction}

The scalar-tensor theory first time was invented by P. Jordan \cite{Jordan}
in the 1950's, and then taken over by C. Brans and R.H. Dicke \cite{Dicke}
some years later. In this theory the scalar field acts as the source of the
(local) gravitational coupling with {\it G }$\sim $ $\chi =\phi ^{-1}$ and
consequently the gravitational constant is not in fact a constant but is
determined by the total matter in the universe through an auxiliary scalar
field equation. In the Jordan conformal frame, the Jordan, Brans, Dicke
(JBD) action takes the form \cite{Jordan} (we use geometrized units such
that G = c = 1 and we follow the signature +,-,-,-).

\begin{equation}
\delta \int \chi ^\eta \left( R-\xi \frac{\chi _{,i}\chi ^{,i}}{\chi ^2}%
+L_m\right) \sqrt{-g}d^4x.  \label{e1}
\end{equation}
where $\eta $ and $\xi $ coupling constants, {\it L}$_m$ is the Lagrangian
density of ordinary matter. Variation of (\ref{e1}) with respect to g$_{ik}$
and $\chi $ gives, respectively, the field equations

\begin{equation}
\begin{tabular}{l}
$\frac{\chi _{,i}^{,i}}\chi +\left( \eta -1\right) \frac{\chi _{,i}\chi ^{,i}%
}{\chi ^2}=\frac{\eta \chi T}{3\eta ^2-2\xi },$ \\ 
\\ 
$R_{ik}+\eta \frac{\chi _{,i;k}}\chi -\left[ \xi -\eta \left( \eta -1\right)
\right] \frac{\chi _{,i}\chi _{,k}}{\chi ^2}=\chi \left( T_{ik}+g_{ik}\frac{%
\xi -\eta ^2}{3\eta ^2-2\xi }T\right) $%
\end{tabular}
\label{e2}
\end{equation}
where R is the Ricci scalar, and {\it T=T}$_{kk}$ is the trace of the matter
energy momentum tensor.

One of the largest concentrations of literature within the area of exact
solution of field equations relativistic gravity theories is static,
spherically symmetric solutions. Some of them are discovered at the early
stage of development of relativistic theories, but up to now they are often
considered as equivalent representations of some ''unique'' solution.
However, even for standard general relativistic spherical compact objects
there are infinitely many exterior solutions of field equations with
spherical symmetry, which describe physically and geometrically different
space-times \cite{Fiziev}. Since Birkhoffs theorem does not hold in the
presence of scalar field, several static solution of the JBD theory is
possible in spherically symmetric vacuum situation \cite{Bhadra}. Many
spacetimes satisfying Einstein's equations, but these are commonly regarded
as non-physical, or unrepresentative of the spacetime we live in. On the
other hand, what makes a solution realistic is somewhat subjective, and will
depend from author to author. Turning now to specific coordinate system,
various interpretations of spherical symmetry have favoured certain
coordinate systems, and hence particular choice for the radial coordinate.
To start with, note that by using the coordinate freedom inherent in general
relativity any static spherically symmetric geometry can be put into a form
where there are only two independent metric components. The most common
coordinate system so common that it has sometimes been referred to as
''canonical'' coordinates - is what we shell term Schwarzschild coordinates.

\begin{equation}
ds^2=-e^\lambda dr^2-r^2\left( d\theta ^2+\sin ^2\theta ^{}d\varphi
^2\right) +e^\nu dt^2.  \label{e3}
\end{equation}

These are defined by taken {\it r} to be the radial coordinate (the radius
of the 2-sphere with proper area 4$\pi ${\it r}$^2$). Schwarzschild
coordinates account for roughly 55\% of the work in general relativity fluid
spheres \cite{Finch}. The second most popular coordinate system used in
fluid spheres work, accounting approximately 35\% of papers, is isotropic
coordinates. In this coordinate system, the spacelike part of metric is
Galilean. Of other coordinate choices used, accounting for the remaining
10\% of papers, the only one to have been frequently used is Polar Gaussian
coordinates \cite{Finch}.

Space structure in Jordan, Brans, Dicke theory is very different from that
in ordinary general relativity. Static, spherically symmetric solutions of
this theory can be better understood by consider in different coordinate
systems. Further we present a few specific examples where we demonstrate
some exact solutions. We do not claim this list is exhaustive.

\section{Static spherically symmetric vacuum solutions of Jordan,
Brans-Dicke theory.}

The first exact solution of JBD field equations were obtained in parametric
form by Heckmann \cite{Heckmann}, soon after Jordan proposed scalar tensor
theory. The first solution describes the geometry of the space-time exterior
to a prefect fluid sphere in hydrostatic equilibrium. One can chose the
static spherically symmetric metric in the form (\ref{e1}) Schwarzschild or
curvature coordinates. Then the solutions of the gravitational field
equations (\ref{e2}) in the vacuum for $\eta $ =1 take the form \cite
{Heckmann} 
\begin{equation}
\begin{tabular}{l}
$r=\frac{r_0}{\sqrt{\tau }\left( \tau ^{-h}-\tau ^h\right) },$ \\ 
\\ 
$e^\lambda =\frac{4h^2}{\left[ \left( \frac 12+h\right) \tau ^h-\left( \frac
12-h\right) \tau ^{-h}\right] ^2},$ \\ 
\\ 
$\QTR{mathit}{e}^\nu =\tau ^{\frac 1B},$ \\ 
\\ 
$\chi =\chi _0\tau ^{\frac{\beta _0}B},$%
\end{tabular}
\label{e4}
\end{equation}
where $\tau $ parameter, r$_0$, $\beta _0$ arbitrary constant and

\[
h^2=\frac 14-\frac A{B^2};^{}A=\frac{\beta _0}2\left( 1+\beta _0\xi \right)
;^{}B=1+2\beta _0. 
\]
From this expression we observe that the solution goes to Schwarzshild
solution as $\beta _0$ $\rightarrow $ 0:

\begin{equation}
\begin{tabular}{l}
$r=r\frac{r_0}{1-\tau },^{}e^\lambda =\frac 1\tau ,^{}e^\nu =\tau ,^{}\chi
=\chi _0$ \\ 
\\ 
$r_0=2m.$%
\end{tabular}
\label{e5}
\end{equation}

On the other hand for {\it A}=0 ($\xi \neq $0, $\xi \neq $2)

\begin{equation}
e^\lambda =\frac 1{1-r_0/r},^{}e^\nu =\left( 1-r_0/r\right) ^{\frac \xi {\xi
-2}},^{}\chi =\left( 1-r_0/r\right) ^{\frac 1{\xi -2}}.  \label{e6}
\end{equation}

($\xi $=0)

\begin{equation}
e^\lambda =\frac 1{1-r_0/r},^{}e^\nu =1,^{}\chi =\sqrt{1-r_0/r}^{}.
\label{e7}
\end{equation}

From (\ref{e4}) it follows that for {\it h}=0 and B$\neq $0

\begin{equation}
r=\frac{r_0}{\sqrt{\tau }\ln \tau },^{}e^\lambda =\frac 1{\left( 1+\frac
12\ln \tau \right) ^2},^{}e^\nu =\tau ^{\frac 1B},^{}\chi =\chi _0\tau ^{%
\frac{\beta _0}B},  \label{e8}
\end{equation}

for ${\it h}$=0 and {\it B}=0 ($\xi $\TEXTsymbol{<}2)

\begin{equation}
e^\lambda =\frac{4r^2}{r_0^2+4r^2},^{}e^\nu =\left( \pm \frac{\sqrt{%
r_0^2+4r^2}-r_0}{2r}\right) ^{\frac 1{\sqrt{-A}}},^{}\chi =\chi _0e^{-\frac
\nu 2}.  \label{e9}
\end{equation}

($\xi $\TEXTsymbol{>}2)

\begin{equation}
e^\lambda =\frac{4r^2}{4r^2-r_0^2},^{}e^\nu =\frac 1{\sqrt{-A}}\arcsin \frac{%
r_0}{2r},^{}\chi =\chi _0e^{-\frac \nu 2}.  \label{e10}
\end{equation}

($\xi $=2)

\begin{equation}
e^\lambda =1,^{}e^\nu =e^{-\frac{2r_0}r},^{}\chi =\chi _0e^{-\frac{r_0}r}.
\label{e11}
\end{equation}

Moreover, there are some special cases. If the {\it h}$^2$= -{\it b}$^2\ $%
\TEXTsymbol{<} 0 then:

\begin{equation}
\begin{tabular}{l}
$r=\frac{r_0}{2\sqrt{\tau }\sin \left( b\ \ln \tau \right) },$ \\ 
\\ 
$e^\lambda =\frac{4b^2}{\left[ 2b\cos \left( b\ln \tau \right) +\sin \left(
b\ln \tau \right) \right] ^2},$ \\ 
\\ 
$\QTR{mathit}{e}^\nu =\tau ^{\frac 1B},$ \\ 
\\ 
$\chi =\chi _0\tau ^{\frac{\beta _0}B},$%
\end{tabular}
\label{e12}
\end{equation}

Note that from solution for $\eta $ =1 we can easy compute \cite{Jordan} the
solution for every value of $\eta $. Usually the most appropriate choice of
variables and coupling constants in which to study this problem is $\phi $
=1/$\chi $, $\eta $ = -1, $\omega $ = -$\xi $. As mentioned before one can
obtain the solution for $\eta $ = -1; under a mere redefinition of scalar
field $\chi $ $\rightarrow $ 1/$\chi $.

For the purpose of solving JBD field equations (\ref{e2}) new and convenient
technique is proposed. This technique will provide significant tool to
handle the sophisticated and intractable problems of JBD fields. For the
further simplification of the problem we will replace variable {\it r} by 
{\it r}($\nu $) \cite{Kozyrev}.

\begin{equation}
r\longrightarrow r\left( \nu \right) .  \label{e12a}
\end{equation}

Hence the solution of the field equations will be assumed to have the form:

\begin{equation}
\begin{tabular}{l}
$r=e^{\frac{\left( 1+2a\right) \nu }2}\sec \gamma ,$ \\ 
\\ 
$e^\lambda =\frac{\kappa ^2}{\left[ \left( 1+2a\right) \cos \gamma -\kappa
\sin \gamma \right] ^2},$ \\ 
\\ 
$\phi =\phi _0e^{a\nu },$%
\end{tabular}
\label{e13}
\end{equation}

where {\it a},{\it \ b}, ${\it \phi }_0$ are arbitrary constants and the
parameters are connected through the constraint

${\it \kappa }${\it \ = }$\sqrt{-1-2a\left( 1+a\left( 2+\omega \right)
\right) },\gamma =\frac{\kappa \left( 2b+\nu \right) }2.$

But there is other spherically symmetric vacuum solution to the JBD field
equations

\begin{equation}
\begin{tabular}{l}
$r=c^{}e^{\frac{\left( 1+2a\right) \nu }2}\sec \gamma ,$ \\ 
\\ 
$e^\lambda =-\frac{\kappa ^4}{-\kappa ^4+\sigma \cos \gamma ^2+k^3\left(
1+2a\right) \sin 2\gamma },$ \\ 
\\ 
$\phi =\phi _0e^{a\nu },$%
\end{tabular}
\label{e13_}
\end{equation}

with $\sigma =2+10a+\left( 28+6\omega \right) a^2+8\left( 5+2\omega \right)
a^3+4\left( 8+6\omega +\omega ^2\right) a^4.$

While the other, known as exterior Brans solutions, corresponds to the
geometry in isotropic coordinates. Four forms of static spherically
symmetric vacuum solution of the JBD theory are constructed by Brans himself 
\cite{Brans}. However, as it has been shown in \cite{Nandi} among the four
different forms of the static spherically symmetric solution of the vacuum
JBD theory of gravity only two classes, Brans class I and class IV
solutions, are really independent; the remaining solutions are their variant.

The Brans class I solution in isotropic coordinates

\begin{equation}
ds^2=-e^\lambda \left( d\rho ^2-\rho ^2d\theta ^2+\rho ^2\sin ^2\theta
^{}d\varphi ^2\right) +e^\nu dt^2.  \label{e14}
\end{equation}
is given by

\begin{equation}
\begin{tabular}{l}
$ds^2=-e^{\lambda _0}\left( 1+\frac B\rho \right) \left( \frac{1-\frac B\rho 
}{1+\frac B\rho }\right) ^{\frac{2\left( \beta -C-1\right) }\beta }\left(
d\rho ^2-\rho ^2d\theta ^2+\rho ^2\sin ^2\theta ^{}d\varphi ^2\right) +$ \\ 
\\ 
$\qquad \qquad +e^{\nu _0}\left( \frac{1-\frac B\rho }{1+\frac B\rho }%
\right) ^{\frac 2\beta }dt^2,$ \\ 
\\ 
$\phi =\phi _0\left( \frac{1-\frac B\rho }{1+\frac B\rho }\right) ^{\frac
C\beta },$%
\end{tabular}
\label{e15}
\end{equation}
with constant condition

\[
\beta ^2=\left( C+1\right) ^2-C\left( 1-\frac{\omega C}2\right) 
\]
were $\nu _0${\it , }$\lambda _0${\it , B, C} are arbitrary constants.

Brans class IV solution can be written as

\begin{equation}
\begin{tabular}{l}
$ds^2=-e^{\frac{\lambda _0+2\left( C+1\right) }{B\rho }}\left( d\rho ^2-\rho
^2d\theta ^2+\rho ^2\sin ^2\theta ^{}d\varphi ^2\right) +e^{\frac{\nu _0-2}{%
B\rho }}dt^2,$ \\ 
\\ 
$\phi =\phi _0e^{\frac C{B\rho }},$%
\end{tabular}
\label{e16}
\end{equation}
with

\[
C=\frac{-1\pm \sqrt{-2\omega -3}}{\omega +2} 
\]

However, the new coordinate representation (\ref{e12a}) simplifies field's
equations. Therefore the coordinate $\nu $ will be used to give a new
solution in the form:

\begin{equation}
\begin{tabular}{l}
$ds^2=-e^{\lambda _0}\frac{\left( \rho ^2-B^2\right) ^2}{4\rho ^4}\left(
d\rho ^2-\rho ^2d\theta ^2+\rho ^2\sin ^2\theta ^{}d\varphi ^2\right) +$ \\ 
\\ 
$\qquad \qquad +e^{\nu _0}\left( -\frac{\rho +B}{\rho -B}\right) ^{2\sqrt{%
\frac 2{2+\omega }}}dt^2,$ \\ 
\\ 
$\phi =\phi _0\left( -\frac{\rho +B}{\rho -B}\right) ^{-\sqrt{\frac
2{2+\omega }}},$%
\end{tabular}
\label{e16a}
\end{equation}

Other examples of exact JBD solution studied in the literature a power
generalization of the Schwarzschild metric \cite{Goswami}, \cite{Campanelli}

\begin{equation}
\begin{tabular}{l}
$ds^2=-A^{n-1}dr^2-A^nr^2\left( d\theta ^2+\sin ^2\theta ^{}d\varphi
^2\right) +A^{m+1}dt^2,$ \\ 
\\ 
$\phi =\phi _0A^{-\frac{m+n}2},$ \\ 
\\ 
$A=1-2\frac{r_0}r,$%
\end{tabular}
\label{e17}
\end{equation}

where {\it m}, {\it n} and {\it r}$_0$ are arbitrary constants. The coupling
constant is found from:

\[
\omega =-2\frac{m^2+n^2mn+m-n}{\left( m+n\right) ^2}. 
\]

The novelty in the current article lies in the fact that we use other
coordinate choices. Polar Gaussian coordinate was first used for general
relativity by Synge \cite{Synge}:

\begin{equation}
ds^2=-dr^2-A^2\left( d\theta ^2+\sin ^2\theta ^{}\right) d\varphi ^2+e^\nu
dt^2.  \label{e18}
\end{equation}

A brief computation leads to

\begin{equation}
\begin{tabular}{l}
$\nu =\frac 1{\sqrt{2\left( 2+\omega \right) }}\arctan \left( a\sqrt{\frac
2{2+\omega }}\left( r+C\right) \right) +b,$ \\ 
\\ 
$A=\sqrt{\left( r+C\right) ^2-\frac{2+\omega }{2a^2}},$ \\ 
\\ 
$\phi =\phi _0e^{-\sqrt{\frac 2{2+\omega }}\arctan \left( a\sqrt{\frac
2{2+\omega }}\left( r+C\right) \right) -b},$%
\end{tabular}
\label{e19}
\end{equation}
where {\it a, b, C} and ${\it \phi }_0$ are arbitrary constants.

On the other hand for $\omega $ = -2

\begin{equation}
\begin{tabular}{l}
$\nu =b-\frac{2a}{2r+C},$ \\ 
\\ 
$A=r+C,$ \\ 
\\ 
$\phi =\phi _0e^{\frac a{r+C}-2b},$%
\end{tabular}
\label{e20}
\end{equation}

Isothermal coordinates defined according to the metric \cite{Synge}

\begin{equation}
ds^2=e^\nu \left( dt^2-dr^2\right) -A^2\left( d\theta ^2+\sin ^2\theta
^{}\right) d\varphi ^2.  \label{e21}
\end{equation}

A brief computation yields

\begin{equation}
\begin{tabular}{l}
$\nu =\frac 1{\sqrt{3+2\omega }}\ln \left( \frac{-2r+2a\sqrt{3+2\omega }+C}{%
2r+2a\sqrt{3+2\omega }-C}\right) +b,$ \\ 
\\ 
$A=-\frac 12\sqrt{-4\left( 2+2\omega \right) a^2+\left( -2r+C\right) ^2}%
\left( \frac{-2r+2a\sqrt{3+2\omega }+C}{2r+2a\sqrt{3+2\omega }-C}\right)
^{\frac 1{2\sqrt{3+2\omega }}},$ \\ 
\\ 
$\phi =\phi _0\frac{2r+2a\sqrt{3+2\omega }-C}{-2r+2a\sqrt{3+2\omega }+C}%
^{\frac 1{\sqrt{3+2\omega }}},$%
\end{tabular}
\label{e22}
\end{equation}

Nikolai Ivanovich Lobacbevsky for the first time in the mathematical
literature, a geometric theory was presented based on all Euclidean
postulates except for the fifth one which is referred to as the Euclidean
postulate of parallels \cite{Lobachevsky}. In the usual way we get for the
three dimensional space the metric form

\begin{equation}
ds^2=d\rho ^2+r^2\left( d\theta ^2+\sin ^2\theta ^{}d\varphi ^2\right) .
\label{e23}
\end{equation}

The difference between the two geometries consists in the dependence of {\it %
r }on $\rho $ : in the Euclidean geometry r = $\rho $ while in the
Lobachevskyan one \cite{Chernikov}

\[
r=\kappa \sinh \left( \frac \rho \kappa \right) 
\]

On the other hand, a radial coordinate of the spatial hyperbolic geometry is
necessary if perfect fluids with cosmological constant are considered \cite
{Bohmer}, this was noted by Weyl already in 1919 \cite{Weyl}. Therefore it
is interesting to analyse solutions to the JBD field equations with
hyperbolic metric.

\begin{equation}
ds^2=-e^\lambda dr^2-\kappa \sinh \left( \frac r\kappa \right) ^2\left(
d\theta ^2+\sin ^2\theta ^{}d\varphi ^2\right) +e^\nu dt^2.  \label{e24}
\end{equation}

If the energy-momentum tensor of matter {\it T}$_{ik}$, vanishes, the system
of equations (\ref{e2}) have the flat vacuum solution. For the case of
hyperbolic metric:

\begin{equation}
e^\lambda =\cosh \left( \frac r\kappa \right) ^2,^{}e^\nu =1,^{}\phi =\phi
_0.  \label{e25}
\end{equation}

It should be noted, however, there is vacuum solution that identical to
solution of general relativity:

\begin{equation}
\begin{tabular}{l}
$e^\lambda =\frac{b\cosh \left( \frac r\kappa \right) ^2\sinh \left( \frac
r\kappa \right) }{-\kappa a+b\sinh \left( \frac r\kappa \right) },$ \\ 
\\ 
$\QTR{mathit}{e}^\nu =\left( b-\kappa ^{}\csc h\left( \frac r\kappa \right)
\right) ,$ \\ 
\\ 
$\phi =\phi _0,$%
\end{tabular}
\label{e26}
\end{equation}

The different possible vacuum solution of equations (\ref{e2}) with
hyperbolic metric yield:

\begin{equation}
\begin{tabular}{l}
$e^\lambda =\frac{b\cosh \left( \frac r\kappa \right) ^2\sinh \left( \frac
r\kappa \right) }{-\kappa a+b\sinh \left( \frac r\kappa \right) },$ \\ 
\\ 
$\QTR{mathit}{e}^\nu =\left( \frac{\left( 2+\omega \right) \left( b-a\kappa
^{}\csc h\left( \frac r\kappa \right) \right) }\omega \right) ^{\frac \omega
{2+\omega }},$ \\ 
\\ 
$\phi =\phi _0\left( \frac{\left( 2+\omega \right) \left( a\kappa \csc
h\left( \frac r\kappa \right) -b\right) }\omega \right) ^{\frac 1{2+\omega
}},$%
\end{tabular}
\label{e27}
\end{equation}

were {\it a, b,} ${\it \phi }_0$ are arbitrary constants.

\section{Conclusion}

When the energy-momentum tensor T=T$_{kk}$ vanishes one can use ${\it \phi }$
= {\it const} for scalar field outside the matter and the solutions of the
JBD theory become the similar as the solutions of the Einstein theory. Then
in empty space there is not a difference between scalar-tensor theories (as
well as vector-metric theories \cite{Bashkov} ) and Einstein theory. In this
case in empty space celestial-mechanical experiments to reveal a difference
between scalar-tensor theories and Einstein theory is not possible.

On the other hand due to highly non-linear character of gravitational
theories, a desirable pre-requisite for studying strong field condition is
to have knowledge of exact solutions of the field equations. Many of those
solutions may be useful for understanding the inherent character of
gravitational theories. In this work we present different classes of the
static spherically symmetric vacuum solution of JBD theory in different
coordinate systems. We found new exact solution in Polar Gaussian,
isothermal and hyperbolic coordinate for the JBD field equations. In closing
we reiterate that while a tremendous amount is already known concerning
static spherically symmetric spacetimes discuss the solutions with
appropriate choice of coordinate systems fruitful for understanding its
physical relevance.

\end{document}